\def\be{\begin{equation}}
\def\ee{\end{equation}}
\def\ba{\begin{array}}
\def\ea{\end{array}}

\documentclass[prl,showpacs,twocolumn,amsmath]{revtex4}
\usepackage{amsfonts}
\usepackage{mathrsfs}
\usepackage{amssymb}
\usepackage{pifont}
\usepackage{epsfig,subfigure,dsfont,amsthm,amsbsy,mathrsfs,amscd}
\def\qed{\leavevmode\unskip\penalty9999 \hbox{}\nobreak\hfill
     \quad\hbox{\leavevmode  \hbox to.77778em{%
               \hfil\vrule   \vbox to.675em%
               {\hrule width.6em\vfil\hrule}\vrule\hfil}}
     \par\vskip3pt}
\usepackage{leftidx}

\input amssym.def
\begin{document}
\title{\large\bf System-environment dynamics of GHZ-like states in noninertial frames}

\author{ Tinggui Zhang$^{1, \dag}$, Hong Yang$^{2}$ and Shao-Ming Fei$^{3}$}
\affiliation{ ${1}$ School of Mathematics and Statistics, Hainan Normal University, Haikou, 571158, China \\
$2$ School of Physics and Electronic Engineering, Hainan Normal University, Haikou, 571158, China \\
$3$ School of Mathematical Sciences, Capital Normal University, Beijing 100048, China \\
$^{\dag}$ Correspondence to tinggui333@163.com}

\bigskip
\bigskip
%\date{2022.10.12}

\begin{abstract} Quantum coherence,
quantum entanglement and quantum nonlocality are important resources
in quantum information precessing.
However, decoherence happens when a quantum system interacts with
the external environments. We study the dynamical evolution of
the three-qubit GHZ-like states in non-inertial frame when one and/or two qubits undergo decoherence. Under the amplitude damping channel we show that the quantum decoherence and the Unruh effect may have quite different influences on the initial state. Moreover,
the genuine tripartite entanglement and the quantum coherence may suffer sudden death
during the evolution. The quantum coherence is most resistent to the quantum decoherence and
the Unruh effect, then comes the quantum entanglement and the quantum nonlocality which is most fragile among the three. The results provide a new research perspective for relativistic quantum informatics.
\end{abstract}

 \maketitle

\section*{Introduction}
The interactions with external environments may result in
energy dissipations or relative phase changes of a quantum system,
leading to the quantum state degenerating from a coherent superposition state
to mixed or single states \cite{whzu}. On the one hand, such decoherence
reduces the quantum advantages of the important resources including
coherence \cite{taml}, quantum entanglement \cite{rpmk} and quantum
nonlocality \cite{ndsv} of the system. On the other hand, the
decoherence also causes the entanglement between the system and the
environment. The dynamics of the system is no longer unitary in this case. It
plays a fundamental role in the
quantum-to-classical transition \cite{decj,masc} and has been
successfully applied in cavity QED \cite{mejx} and ion trap
experiments \cite{cbqc}.

These quantum resources have been investigated mostly in inertial
systems. In 2013 quantum teleportation with a uniformly accelerated
partner has been demonstrated based on quantum entanglement
degenerated in noninertial system \cite{pmgj}. Soon after, Fuentes
et al. \cite{ifrb} observed that entanglement in noninertial frames
is characterized by the observer-dependent properties. Since then,
progresses have been made to the researches on quantum information
theory in noninertial systems
\cite{pmir,jqsj,eljj,jqja,emlj,jwjja,skmk,jwjjb,bnmm,rbtc,ztjja,
sxjl,ztjjb,skjj,ydzy,nfri,xzjj,wgqs,aqjw,smhsb,smzh,jlsj,qczj,smhsc,lcjj,lfxl,smht,jswm,jpka}.

Among the noninertial systems, the Dirac field \cite{eljj,emlj,mgif,djea,smhsc,lfxl}
can be described by a superposition of the Unruh monochromatic modes
from an inertial perspective \cite{wgun},
\begin{equation}\label{zf12}
\begin{array}{l}
|0\rangle=\cos\beta|0\rangle_{I}^{+}|0\rangle_{II}^{-}
+\sin\beta|1\rangle_{I}^{+}|1\rangle_{II}^{-},\\[1mm]
|1\rangle=|1\rangle_{I}^{+}|0\rangle_{II}^{-},
\end{array}
\end{equation}
where $|n\rangle_{I}$ and $|n\rangle_{II}$ are the number states of
the particle outside the (physically accessible) region and the
antiparticle inside the (physically inaccessible) region of the event horizon, respectively. The superscripts $+$ and $-$ denote particle and antiparticle, respectively.
$\cos\beta=(e^{-2\pi \omega c/a})^{-1/2}$, where $a$ is the acceleration
of the observer, $\omega$ is the frequency of the Dirac particle and
$c$ is the speed of light in vacuum. As $\beta$ increases when $a$ increases, in the following the accelerating parameter $\beta$ of the Unruh effect is used instead of $a$.

For two qubit states, the authors in \cite{jwjja} show that the
decoherence and the loss of entanglement resulted from the Unruh
effect will influence each other remarkably. Sudden death of
entanglement may appear for any acceleration when the whole system
undergoes decoherence. However, when only one qubit undergoes
decoherence, such sudden death may only occur when the acceleration
parameter is greater than a critical point. Recently, the genuine
tripartite nonlocality (GTN) and the genuine tripartite entanglement
(GTE) of Dirac fields in the background of a Schwarzschild black
hole for Greenberger-Horne-Zeilinger-like (GHZ-like) states have
been studied \cite{smhsc,lfxl}. It is found that the Hawking
radiation \cite{swha} degrades both the physically accessible GTN
and the physically accessible GTE. The former suffers from sudden
death at some critical Hawking temperature, while the latter
approaches to a nonzero asymptotic value in the limit of infinite
Hawking temperature. Moreover, the Hawking effect cannot generate
the physically inaccessible GTN, but can generate the physically
inaccessible GTE for fermion fields in curved spacetime. More
recently, it is shown that for three-qubit mixed states the Hawking
effect can also generate the physically inaccessible GTN in curved
spacetime \cite{tzxs}.

Recently, the influences of different noisy
environments on quantum coherence and entanglement for W state in
noninertial frames have been investigated
intensively \cite{wzjj,hzhc,smlz,kmou}. Zeng and Cao \cite{hzhc}
studied the evolution and distribution of quantum coherence for
multipartite W and GHZ states of Dirac fields under
amplitude-damping, phase damping and depolarizing channels in the
noninertial frames. Wu et al. \cite{smlz} investigated the quantum
coherence for N-partite W and GHZ states under the local
amplitude-damping environment when $N-1$ observers are accelerated.
They found that quantum coherence is symmetric with
respect to all the observers for GHZ state, but
to two accelerating observers only for W state. In Ref.
\cite{kmou}, the authors studied quantum coherence and entanglement for W
state of Dirac fields under bit flip, phase flip and phase damping
channels in noninertial frames.

In this paper we study the system-environment dynamics for
three-qubit states of the Dirac fields in a noninertial frame. We
consider the most typical amplitude-damping channel \cite{afmm},
which can be modeled by the spontaneous decay of a two-level quantum
state in an electromagnetic field \cite{jmms}. We consider the case
that one and two of the observers move (or stay) in the noisy
environment and investigate whether or not the quantum decoherence
and the loss of the quantumness generated by the Unruh radiation
would influence each other, as well as the sudden death \cite{tyjh}
of coherence, entanglement and nonlocality.

The outline of the paper is as follows. In Section 2 we simply
recall some knowledge about the theory of open quantum systems,
amplitude damping channel and the quantization of the quantumness of three-qubit
X-type states. In Section 3 we investigate the system-environment
dynamics of GHZ-like states in noninertial frames. We summarize
and discuss our conclusions in the last section.

\section*{Some preliminaries}

The evolution of a system state $\rho_S$ of an open quantum systems is governed by \cite{hpfp,hcar,mejx}
$U_{SE}(\rho_S\otimes|0\rangle\langle0|)U_{SE}^{\dag}$,
where $|0\rangle\langle 0|$ represents the initial state of the environment,
$U_{SE}$ is the evolution operator for the system and environment. By tracing over the
environment, one gets the evolution of the system,
\begin{eqnarray}\label{zf3}
L(\rho_S)&=&Tr_E[U_{SE}(\rho_S\otimes|0\rangle\langle0|)U_{SE}^{\dag}]\nonumber\\
&=&\sum_{\mu}\leftidx_{E}\langle\mu|U_{SE}|0\rangle_{E}\ \rho_S \
\leftidx_{E}\langle 0|U_{SE}^{\dag}|\mu\rangle_{E},
\end{eqnarray}
where $|\mu\rangle_{E}$ is the orthogonal basis of the environment,
and $L$ stands for the evolution of the system.
Eq. (\ref{zf3}) can also be expressed as
\begin{eqnarray}\label{zf4}
L(\rho_S)=\sum_{\mu}M_{\mu}\rho_SM_{\mu}^{\dag},\end{eqnarray}
where $M_{\mu}=\leftidx_{E}\langle\mu|U_{SE}|0\rangle_{E}$ are the
Kraus operators \cite{kkra,mdch}. There are at most $d^2$ independent Kraus
operators when the dimension of the system is $d$ \cite{afmm,dwle}.

Consider a three-qubit state $\rho_{ABC}$ in which one subsystem, say $C$, undergoes
the amplitude damping channel. The action of the amplitude damping
channel on the qubit $C$ can be represented by the following
phenomenological map \cite{hpfp,hcar,jtff},
\begin{eqnarray}\label{zf5}
|0\rangle_C|0\rangle_{e_C} \rightarrow |0\rangle_C|0\rangle_{e_C},
\end{eqnarray}
\begin{eqnarray}\label{zf6}|1\rangle_C|0\rangle_{e_C} \rightarrow
\sqrt{1-P}|1\rangle_C|0\rangle_{e_C}+\sqrt{P}|0\rangle_C|1\rangle_{e_C},
\end{eqnarray}
where $|0\rangle_C$ ($|1\rangle_C$) stands for the ground (excited) state of the subsystem $C$, $|0\rangle_{e_C}$ and $|1\rangle_{e_C}$ are the states of the environment with no
and one excitation of its modes, respectively.
Eq. (\ref{zf5}) indicates that the system has no decay and the
environment is untouched. Eq. (\ref{zf6}) shows that the system remains with probability $1-P$ and the environment exits with probability $P$. $P$ is time-dependent,
$P=(1-e^{-\Gamma t})$, where $\Gamma$ is called the decoherence
rate \cite{afmm}. In this paper, we consider the same environment to all the qubits, i.e., $P$ is the same for each subsystem.

Eqs. (\ref{zf5}) and (\ref{zf6}) can also be expressed in the form of Eq. (\ref{zf4}) with Kraus operators \cite{kkra,mdch,jwjja}:
\begin{eqnarray}\label{zf7}M_0=\left(\begin{array}{cc}
    1 &  0  \\
    0 & \sqrt{1-P}
\end{array}\right),\ \ M_1=\left(\begin{array}{cc}
    0 &  \sqrt{P} \\
    0 & 0
\end{array}\right).\end{eqnarray}
That is,
\begin{eqnarray}\label{zf8}L(\rho_C)=\sum_{i=0}^1 M_{i}\rho_CM_{i}^{\dag}.
\end{eqnarray}
When two qubits, say $B$ and $C$, are coupled to the noisy environment independently,
the evolution of the reduced state $\rho_{BC}$ is given by \cite{jwjja}
\begin{eqnarray}\label{zf9}
L(\rho_{BC})=\sum_{i,j=0}^1 (M_{i}\otimes M_j)\rho_{BC}(M_{i}\otimes M_j)^{\dag}.
\end{eqnarray}

In the computational basis $\{ |000\rangle,|001\rangle,\cdots,|111\rangle\}$, the density matrix of a three-qubit X-type state has the following general form,
$$\rho_X=\left(\begin{array}{cccccccc}
    d_1 &  0 & 0 & 0 &  0 & 0 & 0 & f_1 \\
    0 &  d_2 & 0 & 0 &  0 & 0 & f_2 & 0 \\
    0 &  0 & d_3 & 0 &  0 & f_3 & 0 & 0 \\
    0 &  0 & 0 & d_4 &  f_4 & 0 & 0 & 0 \\
    0 &  0 & 0 & f_4^{\ast} &  e_4 & 0 & 0 & 0 \\
    0 &  0 & f_3^{\ast} & 0 &  0 & e_3 & 0 & 0 \\
    0 &  f_2^{\ast} & 0 & 0 &  0 & 0 & e_2 & 0 \\
    f_1^{\ast} &  0 & 0 & 0 &  0 & 0 & 0 & e_1 \\
\end{array}\right).
$$
The GTN of $\rho_X$ is characterized by \cite{kwzj}
\begin{eqnarray}\label{zs10}
S(\rho_X)=\max\{8\sqrt{2}|f_i|,4|N|\},\end{eqnarray} where
$N=d_1-d_2-d_3+d_4-e_4+e_3+e_2-e_1$. According to the Svetlichny
inequality \cite{gsve}, $\rho_X$ is genuine tripartite nonlocal if
$S(\rho_X)>4$.
The GTE of $\rho_X$ is given by \cite{zzsm}
\begin{eqnarray}\label{zs11}
E(\rho_X)=2\max\{0,|f_i|-m_i\}, i=1,2,3,4,
\end{eqnarray}
where $m_i=\sum_{j\neq i}^4\sqrt{d_je_j}$.
The $C_{l_1}$ quantum coherence is given by \cite{taml}
\begin{eqnarray}\label{zs12}
C(\rho_X)=C_{l_1}(\rho_X)=\sum_{i\neq j}|\rho_{ij}|=2\sum_{i=1}^4|f_i|.
\end{eqnarray}

\section*{System-environment dynamics for GHZ-like states}
In this section, let us consider the GHZ-like states
$|GHZ\rangle=\alpha|000\rangle+\sqrt{1-\alpha^2}|111\rangle$ of
the Dirac fields shared by Alice, Bob and Charlie.

Case (i): Alice and Bob stay stationary while Charlie moves with
uniform acceleration. With respect to the Minkowske modes for Alice
and Bob and the Rindler modes for Charlie, by using Eq.
(\ref{zf12}), the GHZ-like states can be written as
$|\psi\rangle_{ABC_{I}C_{II}}=\alpha\cos\beta|0000\rangle
+\alpha\sin\beta|0011\rangle+\sqrt{1-\alpha^2}|1110\rangle$. By
tracing over the inaccessible modes $C_{II}$, we have the following
density matrix,
$$
\rho_{ABC_{I}}=\left(\begin{array}{cccccccc}
    d_1 &  0 & 0 & 0 &  0 & 0 & 0 & f_1 \\
    0 &  d_2 & 0 & 0 &  0 & 0 & 0 & 0 \\
    0 &  0 & 0 & 0 &  0 & 0 & 0 & 0 \\
    0 &  0 & 0 & 0 &  0 & 0 & 0 & 0 \\
    0 &  0 & 0 & 0 & 0 & 0 & 0 & 0 \\
    0 &  0 & 0 & 0 & 0 & 0 & 0 & 0 \\
    0 &  0 & 0 & 0 & 0 & 0 & 0 & 0 \\
    f_1 &  0 & 0 & 0 &  0 & 0 & 0 & e_1 \\
\end{array}\right),
$$
where $d_1=\alpha^2\cos^2\beta$, $d_2=\alpha^2\sin^2\beta$,
$e_1=1-\alpha^2$ and $f_1=\alpha\sqrt{1-\alpha^2}\cos\beta$. Under the bipartition
$AB|C_I$ $\rho_{ABC_{I}}$ can be also expressed as
\begin{eqnarray}&&\rho_{ABC_{I}}\nonumber\\&=&|00\rangle\langle
00|\otimes(d_1|0\rangle\langle 0|+d_2|1\rangle\langle
1|)+|00\rangle\langle 11|\otimes(f_1|0\rangle\langle
1|)\nonumber\\
& &+|11\rangle\langle 00|\otimes(f_1|1\rangle\langle
0|)+|11\rangle\langle 11|\otimes(e_1|1\rangle\langle
1|).\nonumber\end{eqnarray}

Consider that the Charlie's qubit couples to the noisy environment.
From Eqs. (\ref{zf7}) and (\ref{zf8}), $\rho_{ABC_{I}}$ evolves into
\begin{eqnarray}
&&\rho_{ABC_{I}}^{\prime}\nonumber\\&=&\left(\begin{array}{cccccccc}
    d_1+Pd_2 &  0 & 0 & 0 &  0 & 0 & 0 & \sqrt{1-P}f_1 \\
    0 &  (1-P)d_2 & 0 & 0 &  0 & 0 & 0 & 0 \\
    0 &  0 & 0 & 0 &  0 & 0 & 0 & 0 \\
    0 &  0 & 0 & 0 &  0 & 0 & 0 & 0 \\
    0 &  0 & 0 & 0 & 0 & 0 & 0 & 0 \\
    0 &  0 & 0 & 0 & 0 & 0 & 0 & 0 \\
    0 &  0 & 0 & 0 & 0 & 0 & pe_1 & 0 \\
    \sqrt{1-P}f_1 &  0 & 0 & 0 &  0 & 0 & 0 & (1-P)e_1
\end{array}\right).\nonumber\end{eqnarray}
Using Eqs. (\ref{zs10}), (\ref{zs11}) and ((\ref{zs12})), we obtain
\begin{eqnarray}
S(\rho_{ABC_{I}}^{\prime})=&&\max \{8\sqrt{2}\sqrt{1-P}\alpha\sqrt{1-\alpha^2}\cos \beta,\nonumber\\
& &4[\alpha^2\cos^2
\beta+2P\alpha^2\sin^2\beta-\alpha^2\sin^2\beta\nonumber\\&
&+(2P-1)(1-\alpha^2))]\},\nonumber
\end{eqnarray}
\begin{eqnarray}
E(\rho_{ABC_{I}}^{\prime})=&&2\max\{0,\sqrt{1-P}\alpha
\sqrt{1-\alpha^2}\cos\beta-\nonumber\\& & \sqrt{(1-P)\alpha^2\sin^2\beta P(1-\alpha^2)}\},\nonumber
\end{eqnarray}
\begin{eqnarray}C(\rho_{ABC_{I}}^{\prime})=2\sqrt{1-P}
\alpha\sqrt{1-\alpha^2}\cos\beta.\nonumber
\end{eqnarray}

In FIG. 1 we show the behavior of GTN of $\rho_{ABC_{I}}^{\prime}$
for the GHZ state with $\alpha={1}/{\sqrt{2}}$. It is seen that the increase of either
the parameter $\beta$ of the Unruh effect or the decoherence parameter $P$
reduces the GTN. Moreover, the increase of the two parameters will cause the
sudden death of GTN. With the increase of the acceleration
parameter $\beta$, $S(\rho_{ABC_{I}}^{\prime})$ is larger than $4$ at first
and then smaller than $4$, but will not tend to zero. However, with
the increase of $P$ the GTN tends to zero first and then increases.
\begin{figure}[ptb]
\includegraphics[width=0.4\textwidth]{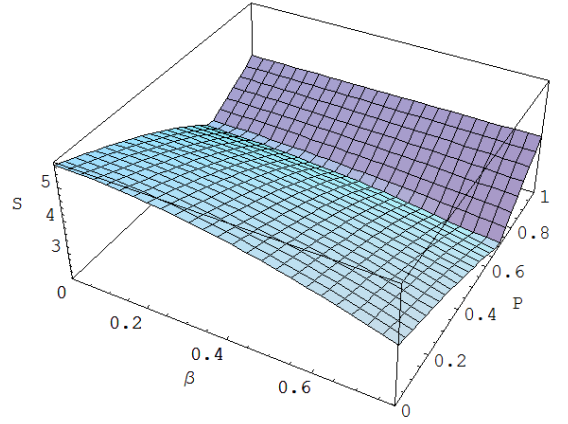}\caption{$S(\rho_{ABC_{I}}^{\prime})$ as a function of the acceleration parameter $\beta$ and decoherence parameter $P$ for the initial GHZ state, when the Charlie's qubit undergoes acceleration and decoherence.}
\label{Fig 1}
\end{figure}

FIG. 2 (a) and (b) show the behavior of GTE and quantum
coherence of $\rho_{ABC_{I}}^{\prime}$ ($\alpha=\frac{1}{\sqrt{2}}$), respectively.
We observe that with the increase of $\beta$ the GTE and quantum coherence
decrease slowly. But the increase of $P$ has a stronger
influence on GTE and quantum coherence, which makes them tend to $0$. And
for large $\beta$, with the decrease of $P$ the GTE and
quantum coherence behavior differently, as GTE is a convex function, while
quantum coherence is a concave function.
\begin{figure}[ptb]
\includegraphics[width=0.4\textwidth]{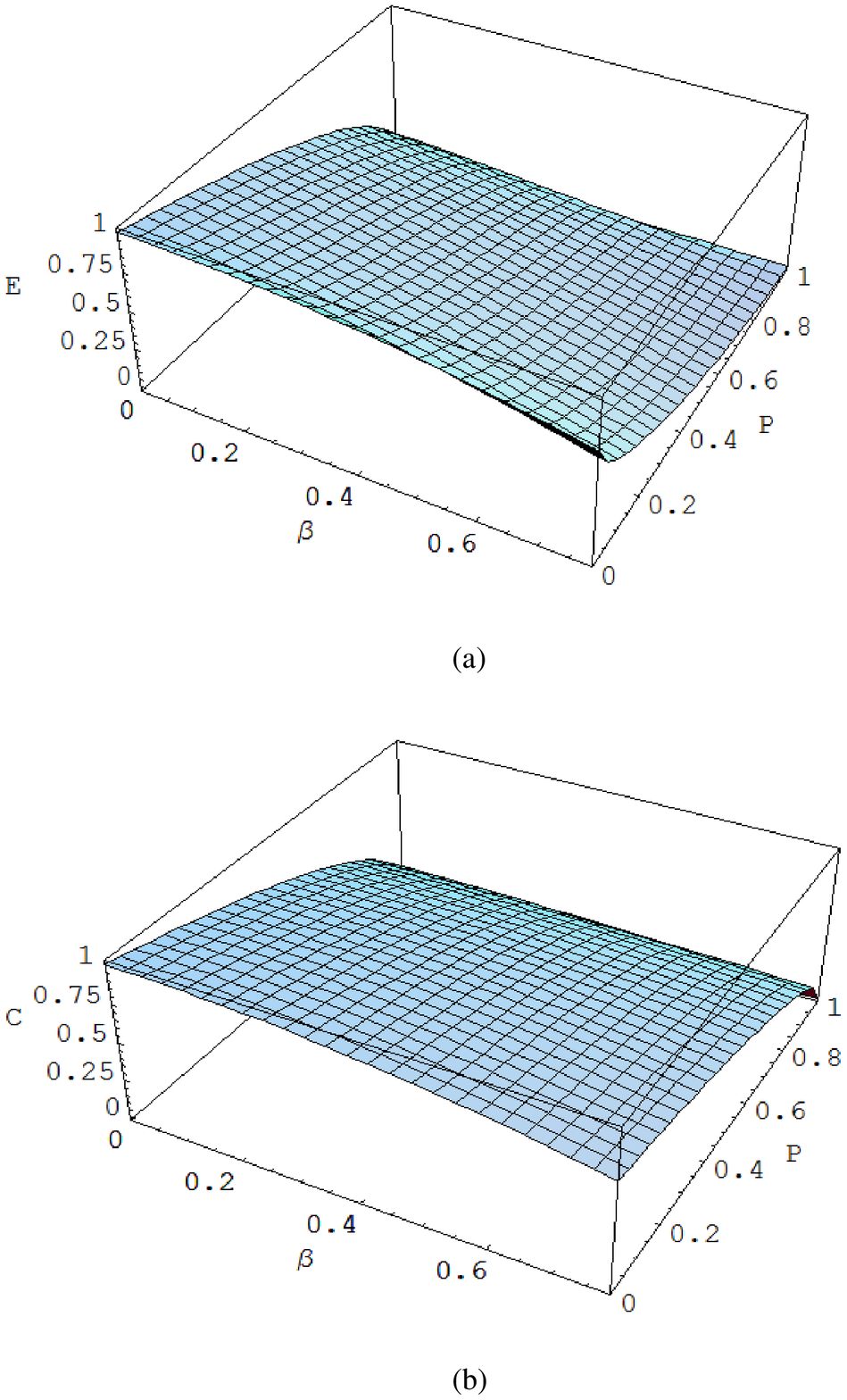}\caption{$E(\rho_{ABC_{I}}^{\prime})$ and $C(\rho_{ABC_{I}}^{\prime})$ as functions of the acceleration parameter $\beta$ and the
decoherence parameter $P$ for the initial GHZ state with $\alpha=\frac{1}{\sqrt{2}}$, when the Charlie's qubit undergoes acceleration and decoherence.} \label{Fig 2}
\end{figure}

Similarly, by tracing over the mode $C_{I}$ we have the reduced state
$\rho_{ABC_{II}}$. Correspondingly, using Eqs. (\ref{zf7}) and (\ref{zf8}) we get
$$\rho_{ABC_{II}}^{\prime}=\left(\begin{array}{cccccccc}
    d_1+Pd_2 &  0 & 0 & 0 &  0 & 0 & 0 & 0 \\
    0 &  (1-P)d_2 & 0 & 0 &  0 & 0 & \sqrt{1-P}f_2 & 0 \\
    0 &  0 & 0 & 0 &  0 & 0 & 0 & 0 \\
    0 &  0 & 0 & 0 &  0 & 0 & 0 & 0 \\
    0 &  0 & 0 & 0 & 0 & 0 & 0 & 0 \\
    0 &  0 & 0 & 0 & 0 & 0 & 0 & 0 \\
    0 &  \sqrt{1-P}f_2 & 0 & 0 & 0 & 0 & e_2 & 0 \\
    0 &  0 & 0 & 0 &  0 & 0 & 0 & 0 \\
\end{array}\right),$$
where $d_1=\alpha^2\cos^2\beta$, $d_2=\alpha^2\sin^2\beta$,
$e_2=1-\alpha^2$ and $f_2=\alpha\sqrt{1-\alpha^2}\sin\beta$.
From straightforward calculation we have
\begin{eqnarray}S(\rho_{ABC_{II}}^{\prime})=&&\max
\{8\sqrt{2}\sqrt{1-P}\alpha\sqrt{1-\alpha^2}\sin \beta,\nonumber\\&
&4[\alpha^2\cos^2 \beta+2P\alpha^2\sin^2\beta+
(1-\alpha^2)-\nonumber\\&
&\alpha^2\sin^2\beta]\}\nonumber\end{eqnarray}
and
\begin{eqnarray}E(\rho_{ABC_{II}}^{\prime})=C(\rho_{ABC_{II}}^{\prime})
=2\sqrt{1-P}\alpha\sqrt{1-\alpha^2}\sin\beta.\nonumber\end{eqnarray}

We plot the GTN of $\rho_{ABC_{II}}^{\prime}$ for GHZ the
state with $\alpha=\frac{1}{\sqrt{2}}$ in FIG. 3(a). It is shown that
the increase of $\beta$ has a greater impact on
$S(\rho_{ABC_{II}}^{\prime})$, while the increase of $P$ has less
impact on $S(\rho_{ABC_{II}}^{\prime})$. However, the values of GTN
are all smaller than $4$. The behavior of GTE (quantum
coherence) of $\rho_{ABC_{II}}^{\prime}$ for the GHZ state
($\alpha=\frac{1}{\sqrt{2}}$) is shown in FIG. 3(b). We see that the
effect of the increase of $\beta$ and $P$ on GTE (or quantum
coherence) is completely the opposite. The GTE (or quantum coherence)
increases with the increase of $\beta$, but decreases with the
increase of $P$. Moreover, the quantum coherence of the initial
state of GHZ-like states
$|GHZ\rangle=\alpha|000\rangle+\sqrt{1-\alpha^2}|111\rangle$
satisfies the following strong nonlinear relationship.
\begin{eqnarray}
C^2(\rho_{ABC_{I}}^{\prime})+C^2(\rho_{ABC_{II}}^{\prime})=4(1-P)\alpha^2(1-\alpha^2),
\end{eqnarray}
which is given by the decoherence parameter $P$, and
has nothing to do with the Unruh effect.
\begin{figure}[ptb]
\includegraphics[width=0.4\textwidth]{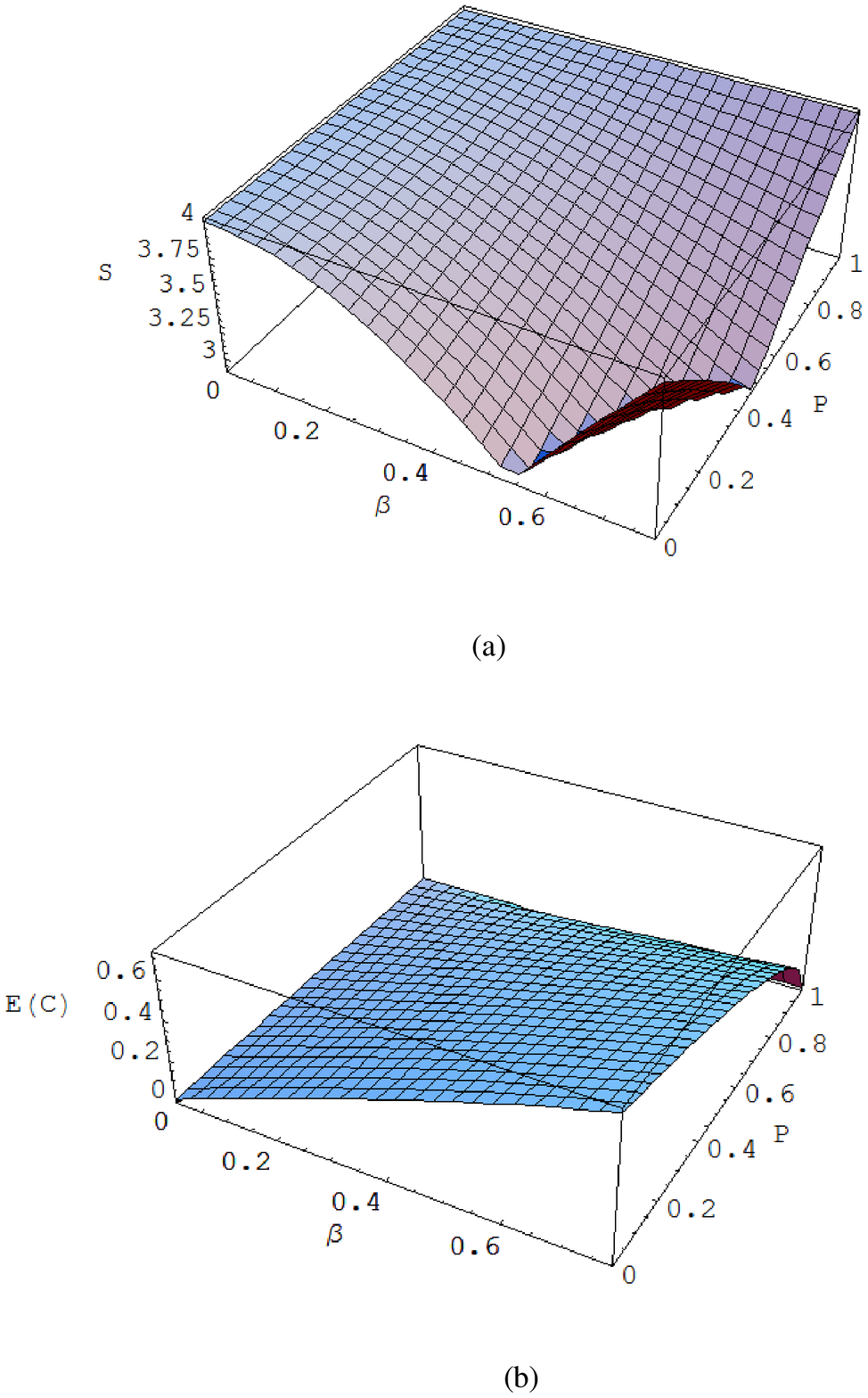}\caption{$S(\rho_{ABC_{II}}^{\prime})$ and $E(\rho_{ABC_{II}}^{\prime})$=$C(\rho_{ABC_{II}}^{\prime})$ as functions
of the acceleration parameter $\beta$ and the decoherence parameter $P$
with respect to the initial GHZ state, when Charlie's
qubit undergoes acceleration and decoherence.} \label{Fig 3}
\end{figure}

Case(ii): we now let Alice still stay at an asymptotically flat
region, while Bob and Charlie move with uniform acceleration $a$.
Using Eqs. (\ref{zf12}), we can rewrite the GHZ-like
states as $|\psi\rangle_{AB_{I}B_{II}C_{I}C_{II}}$, with the detailed
expression given in \cite{smhsc,lfxl}. By tracing over the freedom in the region II, the density matrix $\rho_{AB_{1}C_{I}}$ of the physically accessible part I
is of the form \cite{lfxl},
$$\rho_{AB_{I}C_{I}}=\left(\begin{array}{cccccccc}
    d_1 &  0 & 0 & 0 &  0 & 0 & 0 & f_1 \\
    0 &  d_2 & 0 & 0 &  0 & 0 & 0 & 0 \\
    0 &  0 & d_3 & 0 &  0 & 0 & 0 & 0 \\
    0 &  0 & 0 & d_4 &  0 & 0 & 0 & 0 \\
    0 &  0 & 0 & 0 & 0 & 0 & 0 & 0 \\
    0 &  0 & 0 & 0 & 0 & 0 & 0 & 0 \\
    0 &  0 & 0 & 0 & 0 & 0 & 0 & 0 \\
    f_1 &  0 & 0 & 0 &  0 & 0 & 0 & e_1 \\
\end{array}\right),$$
where $d_1=\alpha^2\cos^4\beta$,
$d_2=d_3=\alpha^2\sin^2\beta\cos^2\beta$, $d_4=\alpha^2\sin^4\beta$,
$e_1=1-\alpha^2$ and $f_1=\alpha\sqrt{1-\alpha^2}\cos^2\beta$. It
can also be written as in the form of bipartition $A|B_{I}C_{I}$,
$$
\rho_{AB_{I}C_{I}}=|0\rangle\langle
0|\otimes N_1+|0\rangle\langle 1|\otimes N_2+|1\rangle\langle
0|\otimes N_3+|1\rangle\langle 1|\otimes N_4,
$$
where
$N_{1}=\left(\begin{array}{cccc}
    d_1 &  0 & 0 & 0  \\
    0 &  d_2 & 0 & 0  \\
    0 &  0 & d_3 & 0  \\
    0 &  0 & 0 & d_4
\end{array}\right),$
$N_{2}=\left(\begin{array}{cccc}
    0 &  0 & 0 & f_1  \\
    0 &  0 & 0 & 0  \\
    0 &  0 & 0 & 0  \\
    0 &  0 & 0 & 0
\end{array}\right),$
$N_{3}=\left(\begin{array}{cccc}
    0 &  0 & 0 & 0  \\
    0 &  0 & 0 & 0  \\
    0 &  0 & 0 & 0  \\
    f_1 &  0 & 0 & 0
\end{array}\right),$
$N_{4}=\left(\begin{array}{cccc}
    0 &  0 & 0 & 0  \\
    0 &  0 & 0 & 0  \\
    0 &  0 & 0 & 0  \\
    0 &  0 & 0 & e_1
\end{array}\right).$

Now we consider that both Bob and Charlie's qubits couple to the noisy
environment independently. From Eqs. (\ref{zf7}) and (\ref{zf9}), state $\rho_{AB_{1}C_{I}}$
evolves to
$$
\rho_{AB_{I}C_{I}}^{\prime}=\left(\begin{array}{cccccccc}
    d_1^{\prime} &  0 & 0 & 0 &  0 & 0 & 0 & f_1^{\prime} \\
    0 &  d_2^{\prime} & 0 & 0 &  0 & 0 & 0 & 0 \\
    0 &  0 & d_3^{\prime} & 0 &  0 & 0 & 0 & 0 \\
    0 &  0 & 0 & d_4^{\prime} &  0 & 0 & 0 & 0 \\
    0 &  0 & 0 & 0 & e_4^{\prime} & 0 & 0 & 0 \\
    0 &  0 & 0 & 0 & 0 & e_3^{\prime} & 0 & 0 \\
    0 &  0 & 0 & 0 & 0 & 0 & 0 & 0 \\
    f_1^{\prime} &  0 & 0 & 0 &  0 & 0 & 0 & e_1^{\prime} \\
\end{array}\right),$$
where $d_1^{\prime}=d_1-P(d_2+d_3)+P^2d_4$,
$d_2^{\prime}=(1-P)d_2+P(1-P)d_4$, $d_3^{\prime}=(1-P)d_3-P(1-P)d_4
$, $d_4^{\prime}=(1-P)^2d_4$, $e_1^{\prime}=(1-P)^2e_1 $,
$e_3^{\prime}=P(1-P)e_1$, $e_4^{\prime}=Pe_1$ and
$f_1^{\prime}=(1-P)f_1$. We obtain
\begin{eqnarray} & & S(\rho_{AB_{I}C_{I}}^{\prime})=\max\{8\sqrt{2}\sqrt{1-P}\alpha\sqrt{1-\alpha^2}\cos
\beta,\nonumber\\&
&4[\alpha^2(\cos^4\beta-2\sin^2\beta\cos^2\beta+(1-2P+2P^2))\nonumber\\&
&-(1-2P+2P^2)(1-\alpha^2)] \}\end{eqnarray}
and
\begin{eqnarray} & &E(\rho_{AB_{I}C_{I}}^{\prime})\nonumber\\& & =2\max\{0, [(1-P)\alpha\sqrt{1-\alpha^2}\cos^2\beta\nonumber\\& &
-\alpha\sin\beta\sqrt{(1-P)\cos^2\beta-(1-P)P\sin^2\beta}\nonumber\\&
& -(1-P)\alpha\sin^2\beta\sqrt{P(1-\alpha^2)}]\}.\end{eqnarray}

We plot the GTN of $\rho_{AB_{I}C_{I}}^{\prime}$ for $\alpha=\frac{1}{\sqrt{2}}$ in FIG. 4(a). It is seen that $S(\rho_{AB_{I}C_{I}}^{\prime})$ is larger than $4$ at first and
then becomes smaller than $4$ with the increase of $\beta$ or $P$. The increase of the two parameters cause the sudden death of GTN. With the increase of acceleration parameter $\beta$,
GTN slowly approaches to $3$, but never reaches $0$.
However, with the increase of the decoherence parameter $P$, the GTN
decreases monotonically and tends to $0$. FIG. 4(b) shows the behavior of
GTE of $\rho_{AB_{I}C_{I}}^{\prime}$ with
$\alpha=\frac{1}{\sqrt{2}}$. One sees that when only the parameter $\beta$
increases, the GTE will not decrease to zero. But when $P$ also exerts
influence, the GTE decreases to zero first, and then increases. Moreover,
compared with FIG. 1 and FIG. 2, we can find that the GTN
and GTE change faster with the change of the parameters. Hence, the decoherence when the two subsystems undergo noisy channel is stronger than that of one system does.
\begin{figure}[ptb]
\includegraphics[width=0.4\textwidth]{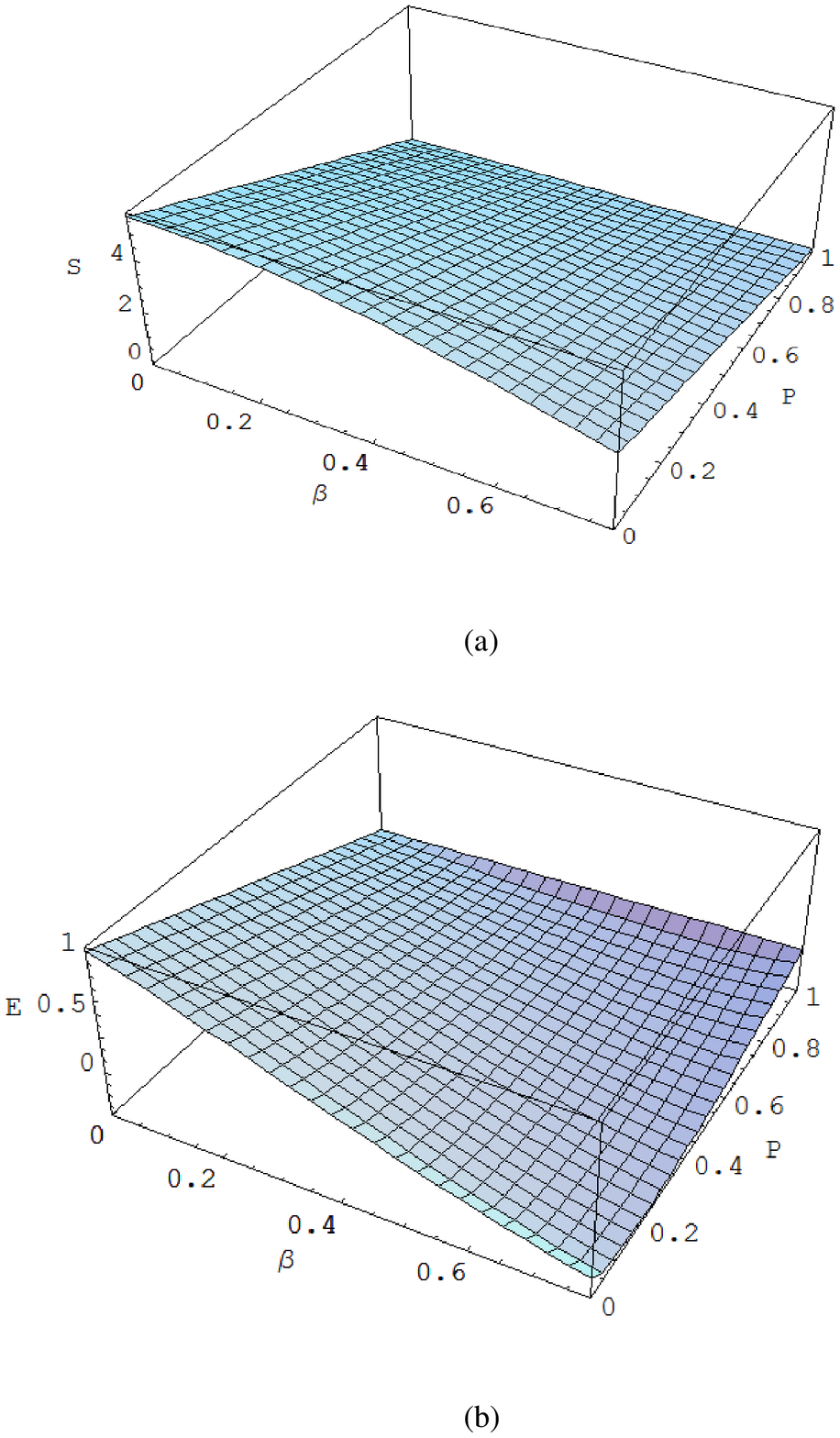}\caption{$S(\rho_{AB_{I}C_{I}}^{\prime})$ and $E(\rho_{AB_{I}C_{I}}^{\prime})$ as functions of the acceleration parameter
$\beta$ and decoherence parameter $P$ for the initial GHZ state,
when Bob's and Charlie's qubits undergo acceleration and decoherence. } \label{Fig 4}
\end{figure}

The quantum dynamics of the rest reduced density matrices can be similarly
analyzed. Due to the symmetry between Bob and Charlie, we only need to consider
the following three situations.
$$
\rho_{AB_{I}C_{II}}^{\prime}=\left(\begin{array}{cccccccc}
    d_1^{\prime} &  0 & 0 & 0 &  0 & 0 & 0 & 0 \\
    0 &  d_2^{\prime} & 0 & 0 &  0 & 0 & f_2^{\prime} & 0 \\
    0 &  0 & d_3^{\prime} & 0 &  0 & 0 & 0 & 0 \\
    0 &  0 & 0 & d_4^{\prime} &  0 & 0 & 0 & 0 \\
    0 &  0 & 0 & 0 & e_4^{\prime} & 0 & 0 & 0 \\
    0 &  0 & 0 & 0 & 0 & 0 & 0 & 0 \\
    0 &  f_2^{\prime} & 0 & 0 & 0 & 0 & e_2^{\prime} & 0 \\
    0 &  0 & 0 & 0 &  0 & 0 & 0 & 0
\end{array}\right),$$
where $d_1^{\prime}=d_1-P(d_2+d_3)+P^2d_4$,
$d_2^{\prime}=(1-P)d_2+P(1-P)d_4$, $d_3^{\prime}=(1-P)d_3-P(1-P)d_4
$, $d_4^{\prime}=(1-P)^2d_4$, $e_4^{\prime}=P(1-\alpha^2)$,
$e_2^{\prime}=(1-P)(1-\alpha^2)$ and
$f_2^{\prime}=(1-P)\alpha\sqrt{1-\alpha^2}\sin\beta\cos\beta$.
According to Eqs. (\ref{zs10}) and (\ref{zs11}), we have
\begin{eqnarray} & &S(\rho_{AB_{I}C_{II}}^{\prime})=\max\{8\sqrt{2}(1-P)
\alpha\sqrt{1-\alpha^2}\sin\beta\cos\beta,\nonumber\\& &4[\alpha^2(\cos^2\beta-\sin^2\beta)^2+(1-2P)(1-\alpha^2)]\},\nonumber
\end{eqnarray}
\begin{eqnarray} & &E(\rho_{AB_{I}C_{II}}^{\prime})=2\max\{0,(1-P)\alpha
\sqrt{1-\alpha^2}\sin\beta\cos\beta\nonumber\\& &-(1-P)\alpha\sin^2\beta\sqrt{P(1-\alpha^2)}\}.\nonumber
\end{eqnarray}
FIG. 5(a) and (b) show the behavior of GTN and GTE of
$\rho_{AB_{I}C_{II}}^{\prime}$ when $\alpha=\frac{1}{\sqrt{2}}$, respectively.
\begin{figure}[ptb]
\includegraphics[width=0.4\textwidth]{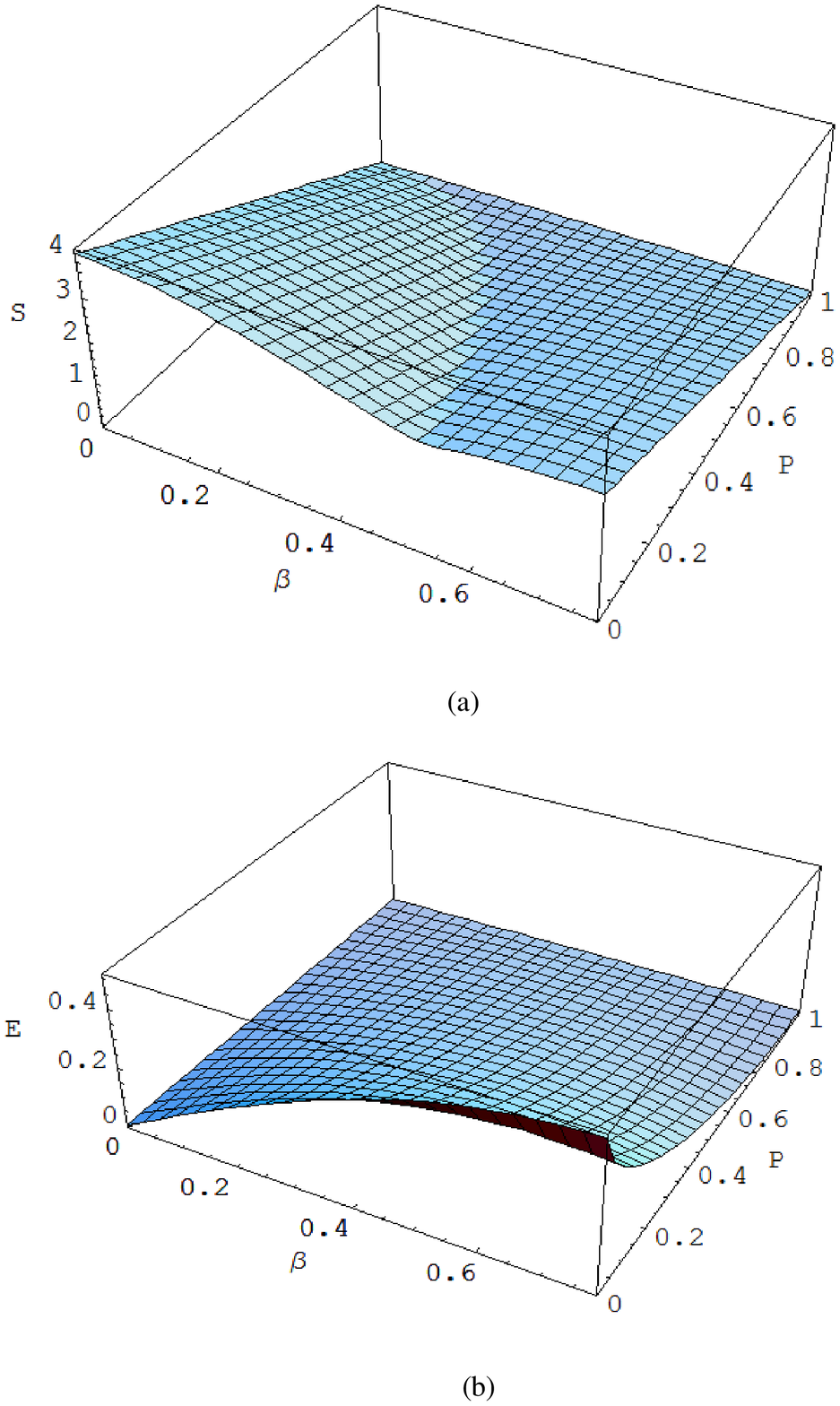}\caption{$S(\rho_{AB_{I}C_{II}}^{\prime})$ and $E(\rho_{AB_{I}C_{II}}^{\prime})$ as functions of the acceleration parameter
$\beta$ and decoherence parameter $P$ for the initial GHZ state, when Bob's and Charlie's qubits undergo acceleration and decoherence.} \label{Fig 4}
\end{figure}

Concerning the systems $A$ and $B_{II}C_{II}$ we have
$$
\rho_{AB_{II}C_{II}}^{\prime}=\left(\begin{array}{cccccccc}
    d_1^{\prime} &  0 & 0 & 0 &  0 & 0 & 0 & 0 \\
    0 &  d_2^{\prime} & 0 & 0 &  0 & 0 & 0 & 0 \\
    0 &  0 & d_3^{\prime} & 0 &  0 & 0 & 0 & 0 \\
    0 &  0 & 0 & d_4^{\prime} & f_4^{\prime} & 0 & 0 & 0 \\
    0 &  0 & 0 & f_4^{\prime} & e_4^{\prime} & 0 & 0 & 0 \\
    0 &  0 & 0 & 0 & 0 & 0 & 0 & 0 \\
    0 &  0 & 0 & 0 & 0 & 0 & 0 & 0 \\
    0 &  0 & 0 & 0 &  0 & 0 & 0 & 0 \\
\end{array}\right),$$
where $d_1^{\prime}=d_1-P(d_2+d_3)+P^2d_4$,
$d_2^{\prime}=(1-P)d_2+P(1-P)d_4$, $d_3^{\prime}=(1-P)d_3-P(1-P)d_4
$, $d_4^{\prime}=(1-P)^2d_4$, $e_4^{\prime}=1-\alpha^2$ and
$f_4^{\prime}=(1-P)\alpha\sqrt{1-\alpha^2}\sin^2\beta$.
According to Eqs. (\ref{zs10}) and (\ref{zs11}), we obtain
\begin{eqnarray} & &S(\rho_{AB_{II}C_{II}}^{\prime})=\max\{8\sqrt{2}(1-P)\alpha
\sqrt{1-\alpha^2}\sin^2\beta,\nonumber\\& &4[\alpha^2(\cos^2\beta-\sin^2\beta)^2-(1-\alpha^2)]\}\nonumber
\end{eqnarray}
and
\begin{eqnarray} & &E(\rho_{AB_{II}C_{II}}^{\prime})=2\max\{0,(1-P)
\alpha\sqrt{1-\alpha^2}\sin^2\beta\}.\nonumber
\end{eqnarray}
In FIG. 6(a) and (b), we plot the GTN and GTE of
$\rho_{AB_{II}C_{II}}^{\prime}$ as functions of $\beta$ and $P$
when $\alpha=\frac{1}{\sqrt{2}}$, respectively.
\begin{figure}[ptb]
\includegraphics[width=0.4\textwidth]{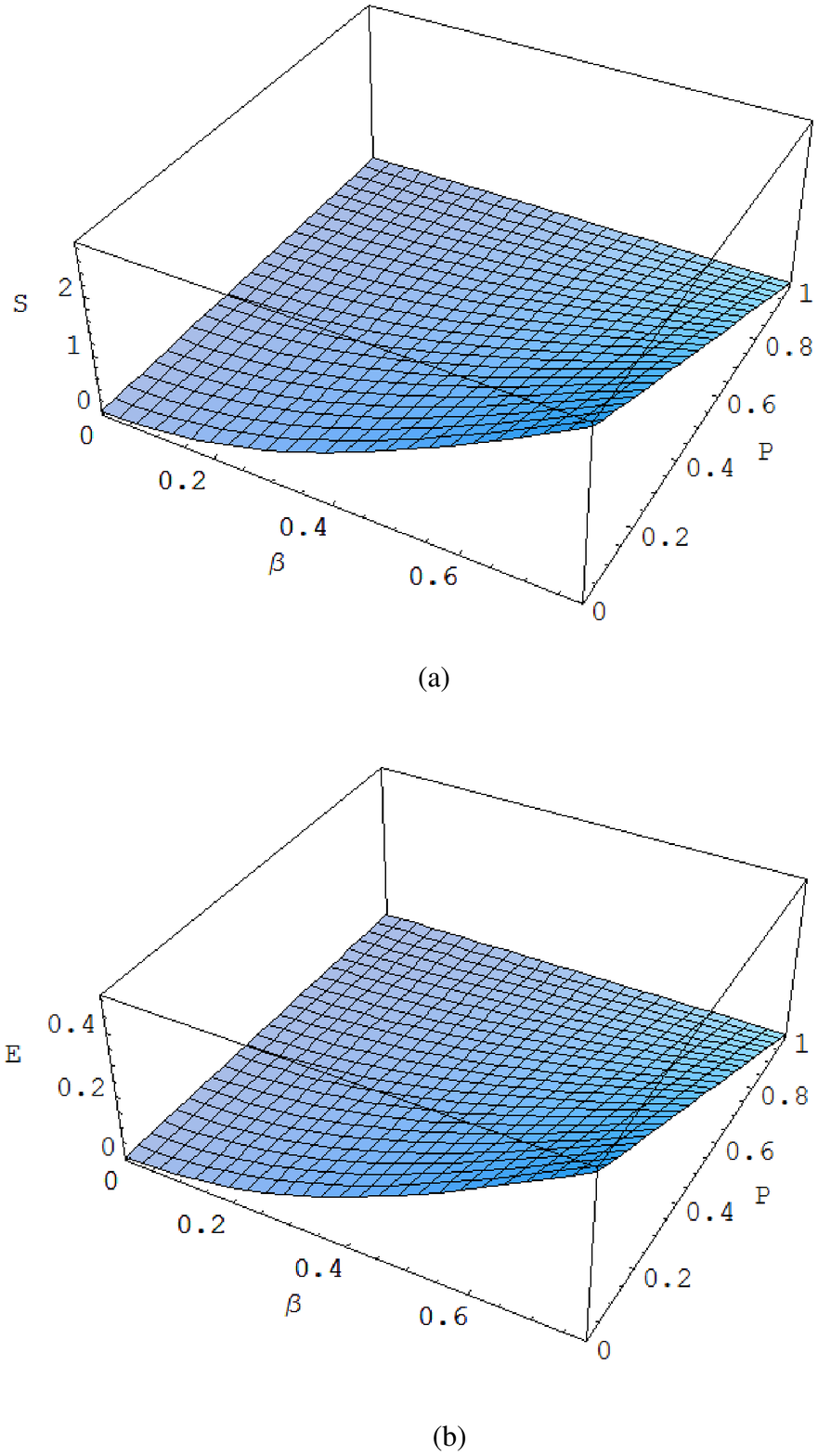}\caption{$S(\rho_{AB_{II}C_{II}}^{\prime})$ and $E(\rho_{AB_{II}C_{II}}^{\prime})$ as functions of the acceleration parameter
$\beta$ and decoherence parameter $P$ for the initial quantum state
is GHZ state, when Bob's and Charlie's qubit undergoes acceleration
and decoherence. } \label{Fig 4}
\end{figure}

With respect to the systems $A$, $B_{I}$ and $B_{II}$, we have
$$\rho_{AB_{I}B_{II}}^{\prime}=\left(\begin{array}{cccccccc}
    d_1^{\prime} &  0 & 0 & 0 &  0 & 0 & 0 & f_1^{\prime} \\
    0 &  0 & 0 & 0 &  0 & 0 & 0 & 0 \\
    0 &  0 & 0 & 0 &  0 & 0 & 0 & 0 \\
    0 &  0 & 0 & 0 & 0 & 0 & 0 & 0 \\
    0 &  0 & 0 & 0 & e_4^{\prime} & 0 & 0 & 0 \\
    0 &  0 & 0 & 0 & 0 & e_3^{\prime} & 0 & 0 \\
    0 &  0 & 0 & 0 & 0 & 0 & e_2^{\prime} & 0 \\
    f_1^{\prime} &  0 & 0 & 0 &  0 & 0 & 0 & e_1^{\prime}
\end{array}\right),$$
where $d_1^{\prime}=\alpha^2\cos^2\beta$,
$e_1^{\prime}=(1-P)^2\alpha^2\sin^2\beta$,
$e_2^{\prime}=P(1-P)\alpha^2\sin^2\beta+(1-P)(1-\alpha^2)$,
$e_3^{\prime}=P(1-P)\alpha^2\sin^2\beta$,
$e_4^{\prime}=P^2\alpha^2\sin^2\beta+P(1-\alpha^2)$ and
$f_1^{\prime}=(1-p)\alpha^2\sin\beta\cos\beta$.
From Eqs. (\ref{zs10}) and (\ref{zs11}) we get
\begin{eqnarray} & &S(\rho_{AB_{I}B_{II}}^{\prime})=\max\{8\sqrt{2}(1-P)
\alpha^2\sin\beta\cos\beta,\nonumber\\& &4[\alpha^2(\cos^2\beta+(2P+2P^2-1)\sin^2\beta)
+(1-P)(1-\alpha^2)]\}\nonumber
\end{eqnarray}
and
\begin{eqnarray} & &E(\rho_{AB_{I}B_{II}}^{\prime})=2\max\{0,(1-P)\alpha^2\sin\beta\cos\beta\}.\nonumber
\end{eqnarray}
In FIG.7 (a) and (b) we plot the behavior of GTN and GTE of
$\rho_{AB_{I}B_{II}}^{\prime}$ for GHZ state
($\alpha=\frac{1}{\sqrt{2}}$), respectively.
\begin{figure}[ptb]
\includegraphics[width=0.4\textwidth]{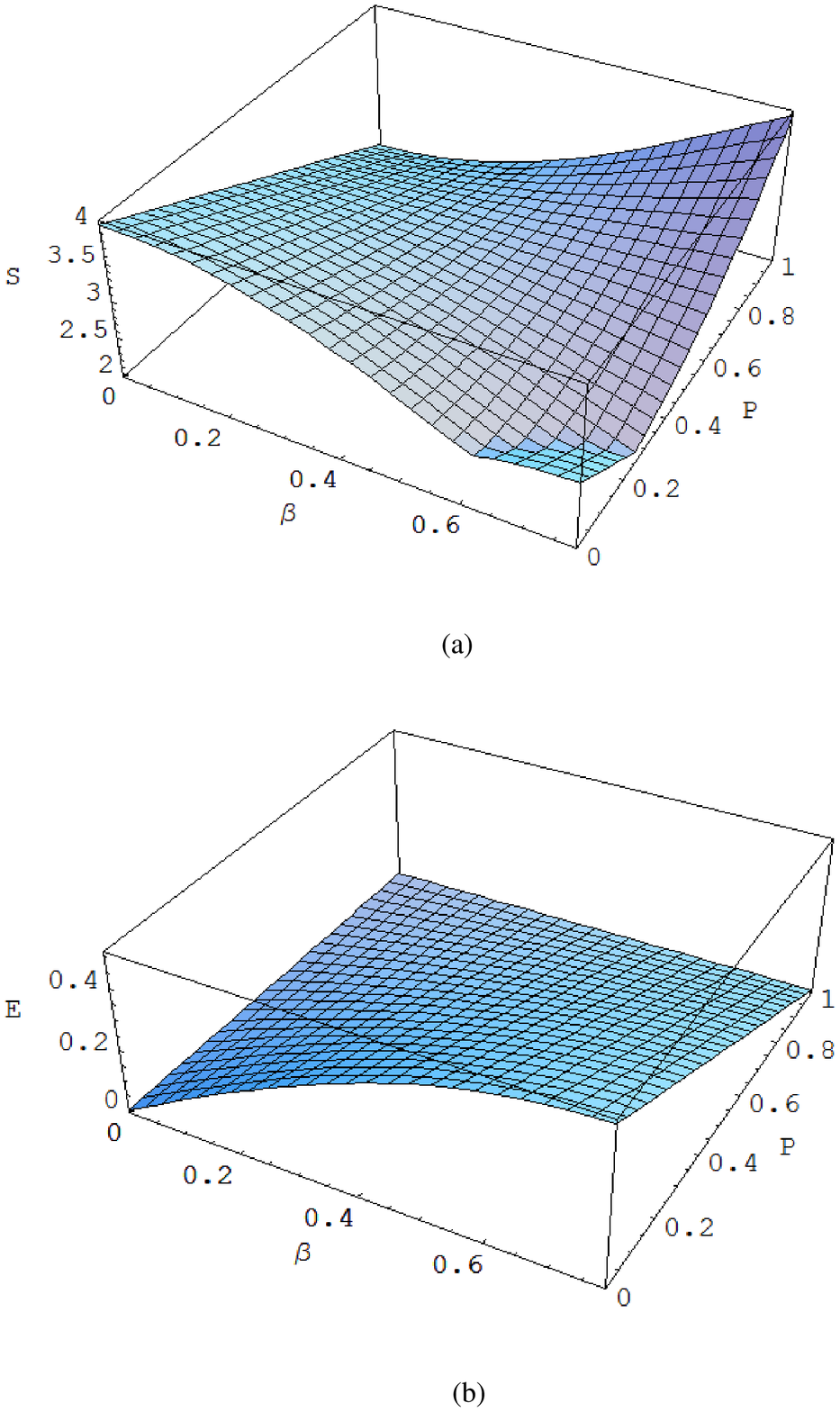}\caption{$S(\rho_{AB_{I}B_{II}}^{\prime})$ and $E(\rho_{AB_{I}B_{II}}^{\prime})$ as functions of the acceleration parameter
$\beta$ and decoherence parameter $P$ when Bob's and Charlie's qubits undergo acceleration
and decoherence.} \label{Fig 4}
\end{figure}

From the figures above, we observe that (1) the influence of the quantum decoherence and the
Unruh effect on the dynamical evolution of the systems are not always in the same rhythms.
(2) For the most cases, the quantum decoherence has a stronger influence than
Unruh effect, as it may result in the phenomena of sudden death. (3) The decoherence of two subsystems is stronger than that of only one subsystem.

In the case (ii), the
influence of $P$ is exerted under the influence of acceleration
parameter $\beta$. We find that throughout the process of
decay the quantum coherence satisfies strictly a beautiful relation.
For example, the following relations always hold,
\begin{eqnarray}\label{zs16}C(\rho_{AB_{I}C_{I}}^{\prime})
+C(\rho_{AB_{II}C_{II}}^{\prime})=2(1-P)\alpha\sqrt{1-\alpha^2},
\end{eqnarray}
\begin{eqnarray}\label{zs17}& & C^2(\rho_{AB_{I}C_{I}}^{\prime})+C^2(\rho_{AB_{II}C_{II}}^{\prime})
+C^2(\rho_{AB_{I}C_{II}}^{\prime})+\nonumber\\& &C^2(\rho_{AB_{II}C_{I}}^{\prime})=4(1-P)^2\alpha^2(1-\alpha^2),
\end{eqnarray}
\begin{eqnarray}\label{zs18}& &C^2(\rho_{AB_{I}C_{I}}^{\prime})+C^2(\rho_{AB_{II}C_{II}}^{\prime})
+(1-\alpha^2)[C^2(\rho_{AB_{I}B_{II}}^{\prime})\nonumber\\& &+C^2(\rho_{AC_{I}C_{II}}^{\prime})]=4(1-P)^2\alpha^2(1-\alpha^2).
\end{eqnarray}
Interestingly, when $P=0$ the right hands of the above
equations are just the quantum coherence of the initial
GHZ-like state. When the decoherence appears,
that is, the value of $P$ increases, the quantum coherence of
the single qubit state decreases at a rate of about
$(1-P)$. Therefore, the larger the $P$ is, the faster the quantum coherence
decreases. When $P\rightarrow 1$, the quantum coherence of a single
reduced density matrix tends to $0$. This phenomena is independent of the
parameter $\beta$ of the Hawking effect.

It is noted that without decoherence ($P=0$), the
entanglement of all the above three-body reduced states is equal to
quantum coherence, namely, Eqs (\ref{zs16}), (\ref{zs17}) and (\ref{zs18})
hold both for entanglement and quantum coherence \cite{smhsc}.
Nevertheless, after decoherence they are valid only for coherence, because
the entanglement of $\rho_{AB_{I}C_{I}}^{\prime}$ and
$\rho_{AB_{I}C_{II}}^{\prime}$ are slightly reduced due to decoherence.
Therefore, the quantum entanglement is more fragile than quantum coherence during the decoherence.

\section*{Conclusions and discussions}
We have studied the dynamical evolution of the three-qubit GHZ-like states in non-inertial frame when one and two qubits undergo decoherence.
Under the amplitude damping channel the influences of
the quantum decoherence and the Unruh effect on the initial states
has been investigated. It is shown that the GTE and quantum coherence may
suffer sudden death. The results can be applied to the cases in which Alice moves
along a geodesic while Bob and/or Charlie hover
near the event horizon with an uniform acceleration.
Our results may also inspire the study on the dynamics of quantum states in the
framework of relativity.
In addition, one can also consider the dynamical behavior
under the influence of amplitude damping or phase
damping for other initial quantum states such as W-state or
mixed states.

\bigskip
{\bf Acknowledgments:} ~This work is supported by the Hainan
Provincial Natural Science Foundation of China under Grant
Nos.121RC539; the National Natural Science Foundation of China
(NSFC) under Grant Nos. 11861031, 12204137, 12075159 and 12171044;
the specific research fund of the Innovation Platform for
Academicians of Hainan Province under Grant No. YSPTZX202215;
Beijing Natural Science Foundation (Grant No. Z190005).

{\bf Data Availability Statement:} This manuscript has no associated
data.


\begin{thebibliography}{99}

\bibitem{whzu} W.H. Zurek, Rev. Mod. Phys. {\bf 75}, 715 (2003).
\bibitem{taml} T. Baumgratz, M. Cramer, M.B. Plenio, Phys. Rev. Lett. {\bf 113},
140401 (2014); \\ A. Streltsov, G. Adesso, M.B. Plenio, Rev. Mod.
Phys. {\bf 89}, 041003 (2017); \\ M.L. Hu, X. Hu, J.C. Wang, Y.
Peng, Y.R. Zhang, H. Fan, Phys. Rep. {\bf 762}, 1 (2018).
\bibitem{rpmk} R. Horodecki, P. Horodecki, M. Horodecki, and K. Horodecki,
Rev. Mod. Phys. {\bf 81}, 865 (2009).
\bibitem{ndsv} N. Brunner, D. Cavalcanti, S. Pironio, V. Scarani and S. Wehner, Rev. Mod. Phys. {\bf 86}, 419 (2014).
\bibitem{decj} D. Giulini, E. Joos, C. Kiefer, J. Kupsch, I.O. Stamatescu, H.D.
Zeh, Decoherence and the Appearance of a Classical World in Quantum
Theory, Springer, 1996.
\bibitem{masc} M.A. Schlosshauer, Decoherence and the Quantum-To-Classical Transition, Springer,
2007.
\bibitem{mejx} M. Brune, E. Hagley, J. Dreyer, X. Maitre, A.
Maali, C. Wunderlich, J.M. Raimond, S. Haroche, Phys. Rev. Lett.
{\bf 77}, 4887 (1996).
\bibitem{cbqc} C.J. Myatt, B.E. King, Q.A. Turchette, C.A. Sackett, D. Kielpinski, W.M. Itano, C. Monroe, D.J. Wineland, Nature
{\bf 403}, 269 (2000).
\bibitem{pmgj} P.M. Alsing, G.J. Milburn, Phys. Rev. Lett. {\bf 91} 180404 (2003).
\bibitem{ifrb} I. Fuentes-Schuller, R.B. Mann, Phys. Rev. Lett. {\bf 95}, 120404 (2005).
\bibitem{pmir} P.M. Alsing, I. Fuentes-Schuller, R.B. Mann, T.E. Tessier, Phys. Rev. A {\bf 74}, 032326 (2006).
\bibitem{jqsj} J. Wang, Q. Pan, S. Chen, J. Jing, Phys. Lett. B {\bf 677}, 186 (2009).
\bibitem{eljj} E. Mart\'in-Mart\'inez, L.J. Garay, J. Le\'on, Phys. Rev. D {\bf 82}, 064006 (2010).
\bibitem{jqja} J. Wang, Q. Pan, J. Jing, Phys. Lett. B {\bf 692}, 202 (2010).
\bibitem{emlj} E. Mart\'in-Mart\'inez, L.J. Garay, J. Le\'on, Phys. Rev. D {\bf 82}, 064028 (2010).
\bibitem{jwjja} J. Wang, J. Jing, Phys. Rev. A {\bf 82}, 032324 (2010).
\bibitem{skmk} S. Khan, M.K. Khan, J. Phys. A Math. Theor. {\bf 44}, 045305 (2011).
\bibitem{jwjjb} J. Wang, J. Jing, Phys. Rev. A {\bf 83}, 022314 (2011).
\bibitem{bnmm} B. Nasr Esfahani, M. Shamirzaie, M. Soltani, Phys. Rev. D {\bf 84}, 025024 (2011).
\bibitem{rbtc} R.B. Mann, T.C. Ralph, Class. Quantum Gravity  {\bf 29}, 220301(2012).
\bibitem{ztjja} Z. Tian, J. Jing, J. High Energy Phys. {\bf 04}, 109 (2013).
\bibitem{sxjl} S. Xu, X.K. Song, J.D. Shi, L. Ye, Phys. Rev. D {\bf 89}, 065022 (2014).
\bibitem{ztjjb} Z. Tian, J. Jing, J. High Energy Phys. {\bf 07}, 089 (2014).
\bibitem{skjj} S. Kanno, J.P. Shock, J. Soda, Phys. Rev. D {\bf 94}, 125014 (2016).
\bibitem{ydzy} Y. Dai, Z. Shen, Y. Shi, Phys. Rev. D {\bf 94}, 025012 (2016).
\bibitem{nfri} N. Friis, New J. Phys. {\bf 18}, 033014 (2016).
\bibitem{xzjj} X. Liu, Z. Tian, J. Wang, J. Jing, Phys. Rev. D {\bf 97}, 105030 (2018).
\bibitem{wgqs} W.C. Qiang, G.H. Sun, Q. Dong, and S.H. Dong, Phys. Rev. A {\bf 98}, 022320 (2018).
\bibitem{aqjw} A.J. Torres-Arenas, Q. Dong, G.H. Sun, W.C. Qiang, S.H. Dong, Phys. Lett. B {\bf 789}, 93 (2019).

\bibitem{smhsb} S.M. Wu, H.S. Zeng, Class. Quantum Gravity {\bf 37}, 115003 (2020).
\bibitem{smzh} S.M. Wu, Z.C. Li and H.S. Zeng, EPL, {\bf 129}, 40002 (2020).
\bibitem{jlsj} J. Wang, L. Zhang, S. Chen, J. Jing, Phys. Lett. B {\bf 802}, 135239 (2020).
\bibitem{qczj} Q. Liu, C. Wen, Z. Tian, J. Jing, J. Wang, Phys. Rev. A {\bf 105}, 062428 (2022).
\bibitem{smhsc} S.M. Wu, H.S. Zeng, Eur. Phys. J. C  {\bf 82}, 4 (2022).
\bibitem{lcjj} L. Xiao, C. Wen, J. Jing, J. Wang, Eur. Phys. J. C {\bf 82},684 (2022).
\bibitem{lfxl} L.J. Li, F. Ming, X.K. Song, L. Ye, D. Wang, Eur. Phys. J. C {\bf 82}, 726 (2022).
\bibitem{smht} S.M. Wu, H.S. Zeng and T. Liu, New J. Phys. {\bf 24}, 073004 (2022).
\bibitem{jswm} J. Jing, S. Long, W. Deng, M. Wang, J. Wang, Sci. China, Phys. Mech. Astron. {\bf 65}, 100411, (2022).
\bibitem{jpka}J. A. Szypulski, P. T. Grochowski, K. Debski and A. Dragan,
arXiv:2112.07250(2021)
\bibitem{mgif} M. Aspachs, G. Adesso, I. Fuentes, Phys. Rev. Lett {\bf 105}, 151301 (2010).
\bibitem{djea} D.E. Bruschi, J. Louko, E. Martn-Martnez, A. Dragan, I. Fuentes,
Phys. Rev. A {\bf 82}, 042332 (2010).
\bibitem{wgun} W.G. Unruh, Phys. Rev. D {\bf 14} 870 (1976).
\bibitem{swha} S.W. Hawking, Nature {\bf 248}, 30 (1974).
\bibitem{tzxs} T. Zhang, X. Wang, S.M. Fei, arXiv:2212.02245 (2022).
\bibitem{afmm} A. Salles, F. de Melo, M. P. Almeida, M. Hor-Meyll, S. P.
Walborn, P. H. SoutoRibeiro, and L. Davidovich, Phys. Rev. A {\bf
78}, 022322 (2008).
\bibitem{jmms} J. M. Raimond, M. Brune, and S. Haroche,
Rev. Mod. Phys. {\bf 73}, 565 (2001).
\bibitem{tyjh}  T. Yu and J. H. Eberly, Phys. Rev. Lett. {\bf 97}, 140403 (2006).
\bibitem{hpfp} H. P. Breuer and F. Petruccione, The Theory of Open
Quantum Systems, Oxford University Press, Oxford, (2002).
\bibitem{hcar} H. Carmichael, An Open Systems Approach to Quantum Optics, Springer,
Berlin, (1993).
\bibitem{kkra}  K. Kraus, States, Effects and Operations: Fundamental Notions of Quantum Theory, Springer, Berlin, (1983).
\bibitem{mdch}  M.D. Choi, Linear Algebr. Appl. {\bf 10}, 285 (1975).

\bibitem{wzjj} W. Zhang and J. Jing, Multipartite entanglement for open system
in noninertial frames. arXiv:1103.4903v1 (2011).

\bibitem{hzhc} H. S.
Zeng and H. M. Cao, Ann. Phys. (Berlin) {\bf 533}, 2000606 (2021).

\bibitem{smlz}  S.M. Wu, Z.C. Li and
H.S. Zeng, Quant. Inf. Process. {\bf 20}, 277 (2021).

\bibitem{kmou} K.I. Kim, M.C. Pak, O.S. An, U.G. Ri, M.C. Ko and N.C.
Kim, Phys. Scr. {\bf 97}, 075101 (2022).

\bibitem{afmm} A. Salles, F. de Melo1, M.P. Almeida1, M. Hor-Meyll, S.P. Walborn, P.H. SoutoRibeiro, L. Davidovich, Phys. Rev. A {\bf 78},  022322 (2008).

\bibitem{dwle} D.W. Leung, J. Math. Phys. {\bf 44}, 528 (2003).
\bibitem{jtff} J. Maziero, T. Werlang, F.F. Fanchini, L.C. Celeri, R.M. Serra, Phys. Rev. A {\bf 81},
022116, (2010).
\bibitem{kwzj} K. Wang, Z.J. Zheng, Sci. Rep. {\bf 10}, 6621 (2020);\\ K. Wang, Y. Liang, Z.J. Zheng, Quantum Inf. Process. {\bf 19}, 140
(2020).
\bibitem{gsve} G. Svetlichny, Phys. Rev. D {\bf 35}, 3066 (1987).
\bibitem{zzsm} Z.H. Ma, Z.H. Chen, J.L. Chen, C. Spengler, A. Gabriel,M. Huber,
Phys. Rev. A {\bf 83}, 062325 (2011); \\ S.M. Hashemi Rafsanjani, M.
Huber, C.J. Broadbent, J.H. Eberly, Phys. Rev. A {\bf 86}, 062303
(2012).

\end{thebibliography}
\end{document}